\documentclass[preprint2]{aastex}

\slugcomment{To be submitted to ApJ}

\shorttitle{Mira's infrared CO wind}
\shortauthors{Ryde et al.}

\begin{document}

\title{Mira's wind explored in scattering infrared CO lines}

\author{N. Ryde, B. Gustafsson and K. Eriksson}
\affil{Uppsala Astronomical Observatory, Box 515, S-751 20, Uppsala, Sweden}
\email{ryde, bg, and ke@astro.uu.se}

\and

\author{K. H. Hinkle}
\affil{National 
Optical Astronomy Observatories\altaffilmark{1}, P.O. Box 26732, Tucson, AZ 85726, U.S.A.}
\email{hinkle@noao.edu}

\altaffiltext{1}{Operated by the Association of Universities for
Research in Astronomy, Inc. under cooperative agreement with the
National Science Foundation}

\begin{abstract}
We have observed the intermediate regions of
the circumstellar envelope of Mira
({\it o}~Ceti) in photospheric light scattered 
by three vibration-rotation transitions of the fundamental band of CO,
from low-excited rotational levels of the ground vibrational state,
at an angular distance
of $\beta\sim2\arcsec\ -7\arcsec\,$ away from the star. 
The data were obtained with the Phoenix spectrometer mounted on the 
4 m Mayall telescope at Kitt Peak. The spatial resolution
is approximately 0.5\arcsec\ and seeing limited. 
Our observations provide absolute fluxes, 
leading to an independent new estimate of the mass-loss rate of 
approximately $3\times 10^{-7}\,\mbox{M$_\odot$ \,yr$^{-1}$}$, as derived from a
simple analytic wind model. 
We find that the scattered intensity from the wind of Mira for 2\arcsec\
{\small
\raisebox{-0.05cm}{\begin{minipage}{0.2cm} 
\raisebox{-0.1cm}{$< $} \\ 
\raisebox{0.1cm}{$\sim$ }
\end{minipage}} }
$\beta$
{\small
\raisebox{-0.05cm}{\begin{minipage}{0.2cm} 
\raisebox{-0.1cm}{$< $} \\ 
\raisebox{0.1cm}{$\sim$ }
\end{minipage}} }
7\arcsec\  
decreases as $\beta^{-3}$, which suggests a time constant
mass-loss rate, when averaged over 100 years, over the past 1200 years.

\end{abstract}


\section {INTRODUCTION}
 
Ever since the circumstellar lines of the $\alpha$ Her M supergiant 
were identified in absorption 
against the photospheric continuum of the 
G III secondary \citep{deutsch}, there have been numerous attempts to map
circumstellar regions. Three spectroscopic tools have been used
extensively.   The most extensively employed are microwave 
emission lines. 
The IR continuum from circumstellar dust, especially in the observations
by the IRAS and ISO satellites, has been used.   
Finally, approximately a dozen 
circumstellar
shells have been imaged in photospheric light scattered by atomic
resonance lines.   
In this paper we shall extend the latter technique
to resonance scattering in molecular lines using the mid-infrared
vibration-rotation lines of the ubiquitous CO molecule.

In his review of
circumstellar envelopes and AGB stars, \citet{habing} noted both the
usefulness and the uniqueness of measuring scattered
resonance lines, for example of Na{\sc i} and K{\sc i}, but drew attention to the
relatively few cases where this method has been applied. This is certainly
because the observations and their interpretation are quite complex.
There are uncertainties in relating these observations to the mass and
structure of the expanding gas since atoms are depleted by dust
grains and the details of ionization in the
circumstellar shell are not known.
An advantage of microwave measurements is that
the molecular species sampled may be much less affected by these
uncertainties. This advantage is also present in the study of
photospheric light scattered by circumstellar CO molecules, as
presented in this paper. The spectral energy distributions (SEDs) of
AGB stars have a maximum in the infrared which also suggests infrared
CO observations. The infrared vibration-rotation 
lines of CO, as an alternative to
rotational CO lines at millimeter wavelengths, also allow the study of
regions close to the star and admit higher spatial resolution in
single telescope studies.

Pioneering investigations of atomic resonance line scattering were made  
of the K{\sc i} 7699\AA\ circumstellar emission in $\alpha$ Ori by
\citet{bernat:75,bernat:76}.  Recent work includes the study of
circumstellar shells around three N-type carbon stars using K{\sc i} 7699
\AA\ \citep{bg:97} and of circumstellar shells of M-type mira stars using
various atomic resonance lines \citep{plez,gui}. 

Imaging off-star emission of CO at $4.6\,\mbox{$\mu$m}$ has previously only been
done a few times \citep{sahai:85,ryde:letter}. \citet{dyck} performed speckle interferometry 
on a number of CO fundamental vibrational-rotational lines. 
Sahai \& Wannier studied the
intermediate regions of the circumstellar envelope of the
dust-enshrouded and bright IR star CW Leonis (IRC+10\,216) by using an
annular aperture, and determined a kinetic temperature of its shell at an average radius of
$2''$. This technique provided only spatially averaged fluxes but
led the authors also to the conclusion that the mass-loss rate in the inner parts
of the circumstellar shell is less than that corresponding to the
region observed in millimeter CO lines.   Since the Sahai and Wannier
observations, great advances have been made in infrared detectors. It
is, however, not trivial to apply infrared arrays in high resolution
spectroscopy since at wavelengths longer than about 1.6 $\mu$m thermal
radiation from room temperature spectrometers dominates the stellar
signals. Cryogenic spectrometers are therefore required, such as the
Phoenix spectrometer \citep{phoenix}. One of the justifications for
building this spectrometer was, in fact, to map circumstellar shells as
presented in this paper.  In \citet{ryde:letter} we demonstrated
methods of observation and analysis of the vibration-rotation CO emission lines
from the intermediate regions of circumstellar winds.  Here, we present
results of an application of these methods to the circumstellar 
envelope of the M4-7IIIe giant {\it o}~Ceti (Mira).

Section 2 discusses Mira. Sections 3 and 4 describe the
observational set-up and the reduction procedures, and the absolute flux calibration, respectively. 
Section 5
reviews the observational results and Section 6 discusses these results in the
light of an analytic model for circumstellar envelopes.

\section{MIRA\label{mira}}

A study of the circumstellar envelope of {\it o}~Ceti (Mira or HD
14386) is of interest for several reasons.
The star  
is the prototype of the
mira-class of 
long-period variables characterized
by large amplitude visual variations.  In the case of {\it o}~Ceti 
the V-band amplitude is more than 6
magnitudes and the mean period is 332 days.
The miras are cool AGB stars with most of the energy emitted in the infrared. 
The brightness of Mira reflects its closeness, 
$(128\pm 18)\,\mbox{pc}$ (ESA 1997\nocite{hipp}).  
The stellar mass loss produces a circumstellar
envelope which, at the distance of Mira, has an angular extension of 2\arcmin\  on the sky 
\citep{loup}.   
Note, however that 
{\it o} Ceti departs from many miras in being  a binary system.
The angular distance between Mira~A and the hot, compact companion star
Mira~B (VZ~Ceti) is 0.6$\arcsec$ (Karovska et al. 
1993; 1997)\nocite{karovska:93}\nocite{karovska:97}.
The companion could be a white dwarf with a mass of about 1\,M$_\odot$, 
a luminosity of 2\,L$_\odot$, and a temperature of more than 
30\,000\,K, which is embedded in an accretion disk, giving rise to 
an abnormal illumination of Mira~A \citep{danchi:94}. Also, note
that 
a number of other high-resolution, optical and infrared wavelength 
measurements provide evidence that Mira itself is elongated 
(see discussion by \citet{ryde:2000} and references therein).

Theoretical model calculations (Bowen 1988; H\"ofner \& Dorfi
1997; H\"ofner et al. 1998\nocite{bowen,dorfi,hoefner:98}; see Woitke
(1998) for a review\nocite{woitke}) indicate that the mass loss of these stars is
most probably caused by a combination of radial pulsation and radiation
pressure on dust grains and/or molecules. It may seem natural to
expect this mass loss to be spherically symmetric, but
stellar rotation may introduce a latitude dependence \citep{dorfi;96}. Also, on the scale
of the spatial resolution of our observations, corresponding to a
time-scale of about 100 years, one could expect the wind to be
homogeneous (observations of Mira show the current pulsational 
behavior dates back to at least 1638). 

However, there are
indications from radio 
observations that the wind of Mira 
departs from this simple picture. 
There is an asymmetry and a shoulder clearly visible in
the radio-line profiles, see for example \citet{planesas:I}. There have been several suggestions in the literature of
different multi-wind scenarios or other phenomena
introducing dramatic variations in the envelope structure with distance
in order to fit the profiles.  
There exist several
combinations of mass-loss rate, expansion and turbulent velocities that are
able to reproduce the Mira circumstellar line profiles rather well
and as a result there is a lack of consensus 
regarding the actual expansion velocities of Mira's wind.
All these results are based on microwave and sub-millimeter observations of 
rotational CO lines.
\citet{crosas:apss} experiment with v$_\mathrm{exp}\approx2$ and
v$_\mathrm{turb}\approx4\,\mbox{km\,s$^{-1}$}$ for their inner wind,
while \citet{young} arrives at v$_\mathrm{exp}=4.8\,\mbox{km\,s$^{-1}$}$
from a fit. However, he points out that the wings show an expansion
velocity of $10\,\mbox{km\,s$^{-1}$}$.  \citet{knapp:98} model the 
circumstellar envelope (CSE)
with a fast outer wind with a mass-loss rate resembling single wind
miras, ${\rm \dot M} = 4.4\cdot 10^{-7}\,\,\mbox{{\rm M$_\odot$
yr$^{-1}$}}$, and an expansion velocity of $(6.7\pm1.0)\,\,\mbox{km\,
s$^{-1}$}$. The inner wind is supposed to be a resumed wind in analogy
with the detached shells found in four carbon stars \citep{ho:96a}.
This slow wind component has a lower mass-loss rate and a much lower
expansion velocity than what is found normally in miras;  ${\rm \dot M}
= 9.4\cdot 10^{-8}\,\,\mbox{{\rm M$_\odot$ yr$^{-1}$}}$ and
v$_\mathrm{exp}=(2.4\pm0.4)\,\,\mbox{km\, s$^{-1}$}$.  Planesas et al.
(1990a) \nocite{planesas:I} interpret their observations of CO($J=1-0$) and CO($J=2-1$) lines
as partly originating from a dominating, spherically symmetric CSE with
v$_\mathrm{exp}= 3\,\mbox{km\,s$^{-1}$}$ and a partial collimation of the 
stellar wind into an additional bipolar lobe. 

Obviously it would be of
great interest to study whether the peculiarities observed in the microwave
lines are also reflected in the infrared CO vibration-rotation lines, which
map the intermediate regions (approximately $100-1000$ stellar radii) of Mira's
circumstellar environment.  

The geocentric, radial
velocity of Mira is large enough for the stellar CO lines to be shifted out of the
telluric lines.  This is a key factor  
since the blending of the
telluric and stellar lines makes the analysis of the faint circumstellar
emission very difficult.  

Multi-wavelength angular diameter measurements of Mira
\citep{haniff:95} combined with Hipparcos trigonometric parallaxes
\citep{leeuwen:97} suggest a mean radius of $(464\pm
80)\,\mbox{R$_\odot$}$.
Based on the work of \citet{mahler:97} on radius and luminosity variations of 
Mira from Wing near-IR photometry, one can derive a bolometric luminosity
$\mbox{L$_{\mathrm{tot}}$}=8900\,\mbox{L$_\odot$}$,
a temperature $\mbox{T$_{\rm{eff}}$}=2400\,\mbox{K}$, and a radius of
$550\,\mbox{R$_\odot$}=3.8\cdot 10^{13}\,\mbox{cm}$ for the phase of
{\it o}~Ceti on the date our observations were made, 1998 October 28-29. 
These values are consistent with the
corresponding values of, for example, \citet{danchi:94} and
\citet{haniff:95}. They arrive at similar temperatures; a mean (over
phases) effective temperature of about T$_{\rm{eff}}\approx2800
\mbox{K}$. We note, however, that this value is low as compared with the values suggested
by \citet{perrin} for giants in the interval M4III-M7III which is
characteristic of miras. Assuming a mass of $ {\cal M }\approx1.0\,\mbox{M$_\odot$}$,
we find the logarithmic surface gravity, $\log g = \log(\mathrm{G\,M/R_*^2)}$, to be 
approximately $-1.0$ [cgs].

\section{OBSERVATIONAL SET-UP AND REDUCTIONS\label{set-up}}

The observations were carried out in 1998 on October 28 and 29 using
the 4 meter Mayall telescope at Kitt Peak with the Phoenix
spectrometer, mounted at the Cassegrain focus of the equatorially
mounted telescope. This cryogenic, single-order echelle spectrometer
is a long-slit, high spectral resolution instrument,
designed for the $1-5\,\mbox{$\mu$m}$ region \citep{phoenix} and marks a major
achievement.  At $4.6\,\mbox{$\mu$m}$ it is now possible to observe objects 
5 magnitudes fainter than was possible with the Fourier Transform
Spectrometer on the same telescope.

The detector is an Aladdin 512$\times$1024 element InSb array cooled to
35 K. The rest of the spectrometer is cooled to 50 K. For our
observations, a $30 \arcsec\,$ long and $0.4 \arcsec\,$ (2 pixels) wide
entrance slit was used, resulting in a spectral resolution of around
$\rm{R}=60\,000$. The slit width projection onto the sky of
0.4\arcsec\,  means that the spectral information is an average over,
at least, a region of this size. The visual seeing was around
$0.8\arcsec\,$, but we note that the seeing disk in the infrared ($4.6\,\mbox{$\mu$m}$) 
is 60\% of that in the visual \citep{lena}, so that the seeing nearly matched the
slit size, which is approximately 30\% greater than the diffraction 
limit of the telescope.
The dispersion of the spectrograph leads to an over-sampled
observed spectrum and each spectrum covers a very small wavelength range;
approximately $11\,\,\mbox{cm$^{-1}$}$.

We observed the circumstellar envelope of {\it o}~Ceti in a hashed
configuration, resembling a perpendicular railway crossing with the
star in the middle, see Figure \ref{planesas}.  The long slit was
placed in the off-star position at $2 \arcsec\,$ away from the star. We
were able to detect emission to a position corresponding to a maximum
distance from the star
of $7\arcsec\,$. The four positions will overlap at
a distance of approximately $\sqrt{2^2+2^2}\arcsec\,$ away from the star, which
enables us to make a relative calibration of the observations, for the
different slit positions around the star.

We chose an echelle setting for observing low excitation lines of the
vibration-rotation, fundamental R-branch of the electronic ground state
of $^{12}$C$^{16}$O.  The lines selected, 1-0 R(1) (2150.86 cm$^{-1}$,
i.e. v$=1\rightarrow 0$ and $J'=2\rightarrow J''=1$), 1-0 R(2) (2154.60
cm$^{-1}$), and 1-0 R(3) (2158.30 cm$^{-1}$), have minimal interference
with telluric lines.  The potentially important 1-0 R(0) line 
has an interfering telluric water line.  Thus, we observed
the region 2150 to 2160 cm$^{-1}$, the spectral coverage being limited
by the array length.  The spectrometer works in order 12
at these wavelengths.

The CO fundamental bands are located in the thermal infrared where the
sky background radiation is high and somewhat variable.  Phoenix is not
a sky balanced device, which means that the thermal radiation is
recorded.  At the $4.6\,\mbox{$\mu$m}$ wavelength of CO, the thermal background 
radiation at Kitt Peak would limit exposure times at $\mathrm{R}=60\,000$ to
about ten minutes.  However, the $4.6\,\mbox{$\mu$m}$ spectrum has a
contribution of telluric lines, mostly CO and H$_2$O, which are seen in 
emission.  To avoid saturation of the telluric lines the exposure time was 
limited to 60 seconds.  

Our observations consist of two types: on-star and off-star exposures. 
On-star exposures are required to remove scattered light from the
off-star exposures.  In order to
remove the thermal spectrum from the source spectrum, 
two exposures are required, one of the source and one of the sky 
\citep{joyce}.  All spectra must also have 
array pixel sensitivity differences removed using flats and darks as
described by \citet{joyce}. 
Since Phoenix has a long slit, nodding along the slit is possible 
for point sources.  Therefore, on-star
spectra were taken by moving 
the star along the slit by $15 \arcsec$ between exposures.  The brightness 
of Mira limited these exposures to $30\,\,\mbox{s}$.  Each set of 
exposures then gave, after differencing, two background-subtracted, on-star
spectra.  Since the stellar signal is extincted by the optical depth 
of the telluric lines, the background-subtracted spectra show in
absorption the 
telluric lines which appear in emission in the raw images. 

Off-star frames were taken the first night by alternatively observing
$2 \arcsec\,$ south and making an identical background observation $62
\arcsec\,$ south of the star, both with an exposure time of
$60\,\,\mbox{s}$. {\it o}~Ceti has a CO shell with a radius of $57
\arcsec$, as measured by radio observations \citep{loup}.  The total
effective observing time in the south position was 66 minutes.
Corresponding observations were subsequently made in the north
position, $2 \arcsec$ N, also here with an exposure time per nod of 60
s. The total effective observing time in the north position was 80
minutes.  The second night a more efficient nodding algorithm was 
employed.  Off-star frames were taken by alternatively
observing $2 \arcsec$ east and $2 \arcsec$ west, 60 s each, then   
a background observation was taken $60 \arcsec$ west off the star.
A total of 54 effective minutes were observed in both the west and east
positions.
Over the few minute (at maximum) interval between the 
source and sky observations 
the sky background level is almost constant
and the airmass changes only slightly.  While the 
thermal radiation and extinction of well-mixed telluric gases, such as 
CO, are scaled with airmass and time, water is not, on most
nights being in the form of clouds visible in the infrared.  As a result the telluric 
water vapor lines will not cancel cleanly in the subsequent differencing of 
spectra.

The on-star spectra were used to remove the
effects of scattered stellar light in the terrestrial atmosphere and in
the spectrometer.  Thus, these were normalized to unit level and then
scaled to the levels of the off-star spectra. The circumstellar CO
vibration-rotation emission was obtained by subtracting the scaled on-star spectra
from the off-star ones.  The emission we wish to detect is about 50
times weaker than the on-star intensity.  The on-star intensity is
typically about twenty to thirty times brighter than the sky
background.

The reduction of the data was performed using standard routines of the
latest version of IRAF (Image Reduction and Analysis Facility). The
wavelength calibration was made by using the telluric lines in every
frame.  The accuracy of the wavelength calibration is of the same order
as the resolution.  Therefore, the uncertainty in the wavelength scale is probably
approximately $0.04\,\mbox{cm$^{-1}$}$ or $6\,\mbox{km\,s$^{-1}$}$.

From the ISO data archive\footnote{Partly based on 
observations with ISO, an ESA
project with instruments funded by ESA Member States 
and with the participation of ISAS 
and NASA.  http://isowww.estec.esa.nl/} we have
retrieved the spectrum around and the flux at $4.6\,\mbox{$\mu$m}$ of Mira, obtained with the Infrared
Space Observatory \citep{kessler}.   The
reductions were made using the most recent pipeline basic reduction
package OLP (v.7) and the ISO Spectral Analysis Package (ISAP v.1.5).
The pipeline processing of the data, such as the flux calibration, is
described by \citet{sws}; the combined absolute and systematic
uncertainties in the fluxes are of the order of $\pm$10\%.

\section{ABSOLUTE FLUX CALIBRATION\label{marcs}}

The absolute calibration of the spectra in the $4\,\mu$m region was
made by a comparison of the measured on-star flux (in counts) with the absolute
flux (in physical units) expected in the broad M-band. The conversion factors
obtained  are then
used for calibrating the off-star spectra. This method contains a number of
uncertain steps, related to the variability of Mira, the uncertainties
in the calibration of broad-band photometry, the scaling from the
broad-band to the narrow wavelength region observed spectroscopically,
the question of how much stellar light was caught within the
spectrograph slit in the on-star observation, and the question whether
the sensitivity of the detector or the transmission of the Earth's
atmosphere might have changed between the on-star and off-star
observations.  In order to check the calibration we therefore 
compared with ISO fluxes and made a
final test in a comparison between the observed spectrum and a calculated
one from a model atmosphere, allowing for the stellar radius and the
distance as measured by the Hipparcos satellite.

The amplitude of the Mira light curve is visually large, but decreases rapidly towards 
longer wavelengths.  From the
NASA Catalog of Infrared Observations \citep{NASA} a mean of eleven
measurements gives a flux in the Johnson M filter of $\log {\cal F_\nu}
= 3.6\pm 0.2$ [Jy].  This M-band mean flux is easily converted to
$(5.6^{+3.3}_{-2.1})\times
10^{-7}\,\mbox{erg\,s$^{-1}$\,cm$^{-2}$\,$\mu$m$^{-1}$}$ or, since 1
pixel corresponds to $0.26\,\mbox{\AA\,}$,
$(1.5^{+0.8}_{-0.5})\times
10^{-11}\,\mbox{erg\,s$^{-1}$\,cm$^{-2}$\,pixel$^{-1}$}$.  In view of
the fact that the flux is peaked towards the blue end of the M-band in such a way that
the flux in the spectroscopic band is higher than the mean flux in the
band (cf. Figure \ref{Mband}), one must correct the absolute calibration.
From the low-resolution ISO spectrum of this region of Mira we find a
correction factor of 1.23. We obtained
this correction by computing a mean of the ISO spectrum over the
M-band and comparing with the ISO flux at $4.64\,\mu$m.
Our model atmosphere (see discussion below) gives a similar result.
Thus, for October 28 we arrive at a conversion factor of
$(3.3^{+1.9}_{-1.2})\times 10^{-14}\,\mbox{erg\,cm$^{-2}$}$ per
detector count and for October 29 we get a factor of
$(2.7^{+1.6}_{-1.0})\times 10^{-14}\,\mbox{erg\,cm$^{-2}$}$ per
detector count. This leads to a flux in the $4.64\,\mbox{$\mu$m}$ band
of $4900\,\mbox{Jy}$ or $6.9\times 10^{-7}\,\mbox{erg\,s$^{-1}$\,cm$^{-2}$\,$\mu$m$^{-1}$}$. 
This may readily be compared with the ISO flux 
at the same wavelength, $(4700\pm 400)\,\mbox{Jy}$, cf. Section \ref{set-up}.

In order to check the calibration further we also used an {\sc os-marcs}
spherical model-atmosphere 
for the stellar parameters 
T$_\mathrm{eff}=2400\,\mbox{K}$, $\log g=-1.0$ [cgs], and solar
metallicity and allowing for the stellar radius (R$_*=3.8\cdot
10^{13}\,\mbox{cm}$) and a Hipparcos distance $\mathrm{d}=128\,\mbox{pc}$, to
generate a synthetic spectrum of the region of interest. 
The stellar parameters are determined for the phase 
of {\it o} Ceti at October 29, 1998, the date the NIR observations were made, see \citet{ryde:2000}.
The model atmosphere is a
part of a new grid of model atmospheres (Plez et al. 2000)
being generated by an extensive update of the {\sc marcs}
code and its input data (based on Gustafsson et al.
1975\nocite{marcs:75}). Molecular lines of H$_2$O (with line lists of
Partridge and Schwencke 1997)\nocite{par}, of CO (lists of Goorvitch
1994)\nocite{goor}, and of many other molecules, are taken into account
in the spherical radiative transfer for the calculation of a synthetic
spectrum. 

Figure \ref{SED} shows the spectrum generated from
the model atmosphere and a comparison with the observed one. The general 
features are well reproduced
and are mainly due to photospheric water vapor and CO. The agreement
of the over-all flux level is astonishingly good; the model flux is
around 30\% too small. We find this agreement between the empirical, calibrated, absolute
flux and the flux we calculate from a model atmosphere
very satisfactory, in view of all the uncertainties anticipated.

It is worth noting that synthetic spectra from the new generation
of spherical {\sc marcs} models in the near infrared are able to 
reproduce observed spectra fairly well, even for a pulsating star like Mira.
While the cyclic variations in mira spectra are well known and
conspicuous at high resolution (Hinkle et al. 1982; Hinkle et al.
1984), little is know about phase dependent variations of the 4.6
$\mu$m photospheric spectra of mira stars.  It is possible that depths
of formation of the continuum and lines at $4.6\,\mu$m are in a
region of the photosphere similar to that of the visual spectrum where
the dynamic behavior does not have large effects on absorption line
formation.  However, it is also possible that the phase of observation
gave a fortuitous match to the model.  In either case the stellar 
atmosphere is extended and the {\sc marcs} match to the spectrum is 
impressive.

Figure \ref{hastighet} shows a blow-up of the photospheric CO R(2)
absorption line as well as the scattered circumstellar R(2) emission. 
This emission line is shifted by $0.42\,\mbox{cm$^{-1}$}$ compared to the telluric 
wavelength scale,
corresponding to a radial velocity of $58.6\,\mbox{km\,s$^{-1}$}$.
Mira has a
v$_\mathrm{LSR}=47\,\mbox{km\,s$^{-1}$}$ deduced from radio data
\citep{loup}, which equals a heliocentric velocity of
$57\,\mbox{km\,s$^{-1}$}$.  At the time of the observations and at Kitt
Peak, this corresponds to a geocentric radial velocity of
$59\,\mbox{km\,s$^{-1}$}$, in excellent agreement with the observations.
Thus, the wavelength of the scattered light is
centered on the laboratory wavelength corrected for the stellar radial
velocity. The photospheric absorption lines, however, are also shifted due to the
velocity of the pulsating photosphere.
The bisector of the photospheric absorption line lies less than $2\,\mbox{km\,s$^{-1}$}$ 
blue-wards of the center of the emission line.
The weak emission in the absorption line in the on-star spectrum is
due to scattered light from the CSE. This light is included in the
on-star measurement since the long-slit will also cover the CSE in two
directions away from the star. Also present in the on-star spectrum 
is additional absorption along the line-of-sight through the
circumstellar shell.  In our measured, off-star, scattered CO
light we do not correct for these components.  However, they are of 
nearly equal intensity and should cancel at the level of the
uncertainties of our measurements.  

Note that when discussing relative fluxes, such as in Section \ref{obs} where 
the R(1)/R(3) and R(2)/R(3) ratios are calculated, the details in the flat-fielding may
introduce uncertainties. The three lines
span over the entire range of the detector, and therefore variable effects not taken care of
by the flat-fielding will show up as an error in the ratios.

\section{OBSERVATIONAL RESULTS\label{obs}}

Figure \ref{on_off} shows the off-star spectrum of the west position. 
Superimposed is the corresponding, scaled, on-star spectrum, which
at least approximately represents the radiation exciting the
molecules.  The intensity of this spectrum is scaled in order to fit the general
features in the off-star spectrum. The off-star spectrum obviously
consists partly of stellar light that is scattered in the Earth's
atmosphere, in the telescope, the spectrometer, and/or by dust grains
in the circumstellar shell, leading to a spectrum resembling the
on-star one.  Nearly all features are identified as photospheric CO and
H$_2$O lines, cf. Section  \ref{marcs} and Figure  \ref{SED}.  The on-star
spectrum includes tens of CO vibration-rotation lines of various
excited vibrational states. The cold off-star spectrum, however, 
includes only the 1-0 R(1), R(2), and R(3) vibration-rotation emission lines
of $^{12}$CO within the observed spectral range, except for the
scattered, on-star light.   Thus, the circumstellar molecules are radiatively excited by the
stellar light, which is re-emitted as emission.

Figure \ref{em} shows the resulting CO emission from the observations
west, east, north and south of {\it o}~Ceti, integrated over the
long slit. This emission is recovered from the data by subtracting the 
scaled on-star
spectrum from the off-star one. The resulting emission lines are the
circumstellar R(1), R(2) and R(3) lines. Variations in the telluric
lines during the time between the object frame exposure and the
background exposure and/or during the time between the on-star and
off-star exposures would result in non-zero residuals.  From the
amplitudes of the signals of the CO emission lines, measured at different directions 
from the star, and the noise level as
shown in Figure \ref{SN}, we estimate a signal-to-noise ratio of approximately 
$5-15$.
The noise arises partly from spurious mismatches in intensity 
between the on- and off-star spectra.

To study the emission as a function of the angular distance from the
star, we divided our long-slit spectra into 79 sub-spectra (symmetrically 
around the maximum intensity representing the closest point to the star), one
spectrum per pixel in the spatial direction. In this way we obtained 79
spectra for every off-star slit position, each corresponding to
0.2\arcsec\ on the sky, if seeing is neglected. 
For every spectrum, the
intensity of the three CO vibration-rotation emission lines [R(1), R(2) and R(3)] from
the wind are measured.  These spectra provide data representing a distance
range from $2\arcsec\,$ to a maximum of $7\arcsec\,$ away from the
star, every slit position giving two series of data with three 
line-fluxes per spectrum, see Figures \ref{beta3}, 
\ref{beta3_R1}, \ref{beta3_R2}, and \ref{beta3_R3};
for example, the west position will sample the south-west and the 
north-west regions of the
wind.  The two sequences in every panel in
the Figures represent these two series.  In Figure \ref{beta3} the decline
of {\it the added} R(1), R(2) and R(3) intensities for the four slit
positions are shown.

Table \ref{kvoter} gives the observed intensities of the CO emission lines
from the spectra representing the closest points to the star, i.e.
$(2\pm 0.5)\arcsec\,$ away from the star,  as well as the mean
intensity ratios, R(1)/R(3) and R(2)/R(3), for positions from
2.0$\arcsec$ to 3.4$\arcsec$. 
Since the S/N ratio at the observed spatial resolution
is too low, the data will unfortunately not permit a useful
plot of the intensity ratios as a function of distance;  The scatter
is too large. 
The best-fit slopes of the intensity as
a function of angular distance from the star, d$\log$~I/d$\log \beta$,
for the R(2) emission line are also given in the Table.  The
uncertainties quoted are pure measuring uncertainties, i.e. they do not 
include possible
systematic uncertainties.  We obtained the best signal for our west position.
The intensity decreases by a factor of about 40 from $2\arcsec\,$ to
$7\arcsec$.  A mean slope of the west, east, and north positions is
$\mathrm d\log I /\mathrm d\log \beta = -2.8\pm 0.3$. Here we omit the south
measurement in order to lower the uncertainty.  Naturally, the
signal-to-noise ratios decrease rapidly outwards.  The uncertainty quoted
is the measuring uncertainty.  

The R(1) emission line is situated close to the left (red) edge
of the detector. The spectra show a peculiar rise in intensity here
which makes the measured values of the R(1) lines subject to a
systematic uncertainty.  Especially the east measurement of the R(1) line
seems to be stronger than expected when compared with the measurements
in the other directions. The R(1)/R(3) ratios are also affected by
this effect.  The east and north spectra in Figure \ref{em} show a
spurious decline at lower wavenumbers (at the red end). This is due to a 
mismatch in intensity between the on- and off-star spectra at the edge of the
detector array probably not caused by flat-fielding or other reduction procedures. 
This spurious effect may, unfortunately, affect all frames.

\section{DISCUSSION}

We find a
power-law dependence of the intensity as a function of angular distance
on the sky, $\mbox{I}\propto \beta^{-3}$.
Furthermore, in view of various uncertainties, e.g. in the positioning of the slit,
we find that the measured emission line intensities in different
directions from the star are consistent with a symmetric wind.
The wind is at least symmetric to within a factor of two in density.
A true west and east
position of $1.8\arcsec\,\mathrm{W}$ and $2.2\arcsec\,\mathrm{E}$ from the star would yield
our measured values for a symmetric wind. Corresponding values for the
north and south positions would be $1.7\arcsec\,\mathrm{N}$ and $2.3\arcsec\,\mathrm{S}$.
Thus, a small error in the position of $0.2-0.3\arcsec$ (which is approximately 
the accuracy of the positioning of the slit) in the
measurement from a symmetric wind would yield the `asymmetric' values
we measured.

The relative intensities measured on the two nights can be checked at the
cross-over points, cf. for example Figure \ref{beta3}.  
The intensities differ by approximately 50\%, probably reflecting the uncertainties
in the slit positions, but also the accuracy of the absolute 
flux calibration between the nights.
The flux calibrations of the observations on the two nights are based on two different 
on-star measurements, 
which are subject to
different uncertainties, as discussed in Section\,\ref{marcs}. 


However, a comparison of the two direction in one slit position may
reveal asymmetries in the wind. In Figures \ref{beta3_R1}, 
\ref{beta3_R2}, and \ref{beta3_R3}, which represent the three lines measured,
the increase in intensity at 3-4\arcsec\ in the `north-east' data in the 
`north'-panel
is seen in all transitions. This is also the case for the 
`south-east' data in the `south'-panel which all show approximately the same 
morphology. 
Thus, this could indicate that there is an east-west asymmetry in 
the intensity measured, reflecting an asymmetric distribution of CO.
This could be due to the additional bipolar 
outflow detected at larger scales in the 
aperture synthesis CO maps by
\citet{planesas:II}. Note that their highest resolution is approximately 6\arcsec.


We now make a simple analytic model of our observations by assuming a spherically 
symmetric and
homogeneous wind with a constant
mass-loss rate and a constant expansion velocity. 
We also assume the wind to be optically thin in the CO lines along the line-of-sight
for rays from regions beyond a certain distance from the star.
The adequacy of this latter assumption will be investigated below.
Based on Eq.\,(5) by \citet{bg:97} the ratio of the
wavelength-integrated, line-scattered intensity, ${\rm I_{CO,i}}$
(erg\,s$^{-1}$\,cm$^{-2}$\,arcseconds$^{-2}$), and the line-scattering
flux $\bar {\rm f}_{\lambda}$ (erg\,s$^{-1}$\,cm$^{-2}$\,cm$^{-1}$) as
seen by the scattering molecules averaged across the line width but
measured at the distance d, is found to be

\begin{eqnarray}
\label{bg_formel}
\frac {{\rm I_{CO,i}}(\beta)}{\bar {\rm f}_{\lambda}} & = & \frac{206265}{32}\,    
\frac{e^2 \,\lambda ^2}{m_e \,c^2\,m_H}\, f_{u\leftarrow l}\, \dot {\rm M} \times \, \nonumber \\
& & \frac{{\mathrm{N_i(CO)}/\mathrm{N(CO)}}\cdot \epsilon_{\mathrm{CO}}}{\mu\,v_e \,{\rm d}} 
\left (\frac{1}{\beta}\right )^3,
\end{eqnarray}

\noindent
where $f_{u\leftarrow l}$ is the absorption oscillator strength of the
line, $\epsilon_{\mathrm{CO}}$ is the fractional abundance of CO
molecules, i.e. [CO]/[H], and N$_i$(CO) denotes the number density of CO molecules in
the lower state, i, of the transition. Furthermore, $\mu$ is the mean
molecular weight, d is the distance to the star, $v_{\mathrm e}$ is the
terminal expansion wind velocity and $\beta$ is the angular distance
from the star on the sky in seconds of arc.  A feature of this general
expression is the minus third power dependence of the scattered intensity
as a function of the impact parameter on the sky, $\beta$, which agrees within the uncertainties
with our observation (cf. Figure \ref{beta3}). Eq.\,(1) will now be
applied to the data for the R(2) line for which our data have the
highest quality.

The oscillator strength $f_{u\leftarrow l}$ of the R(2)
vibration-rotation CO-line at 4.6~$\mu$m is $6.1\times 10^{-6}$
(Kirby-Docken and Liu 1978\nocite{kirby} provide oscillator strengths,
which for our transitions are consistent with Goorvitch \& 
Chackerian\nocite{goor_chack} (1994) and
Hur\'e \& Roueff (1996)\nocite{hure}\footnote{Note the missing cube in their Eq.\,(3)}. See
also Table \ref{einstein}).  Thus, from Eq.\,(\ref{bg_formel}) we find
that the mass-loss rate as deduced from the emitted intensity from the
R(2) line of CO, and, assuming $\mu\approx1.2$, is

\begin{eqnarray} \label{mdot}
 \dot {\rm M} & = & 4.20\times 10^{-10}\cdot \frac{v_e \,{\rm
 d}\,\beta^3}{\mathrm{N_i(CO)}/\mathrm{N(CO)} \cdot
 \epsilon_{\mathrm{CO}}}  \times \nonumber \\
   & & \frac {{\rm I_{CO,i}}(\beta)}{\bar {\rm
   f}_{\lambda}}\,\,\,\,\,\,(\mathrm{M_\odot \,yr^{-1}}),
\end{eqnarray}

\noindent
where $v_\mathrm{e}$ is given in \mbox{km\, s$^{-1}$}, d in
parsecs, $\beta$ in seconds of arc, ${\rm I_{CO,i}}$  in
erg\,s$^{-1}$\,cm$^{-2}$\,arcseconds$^{-2}$, and $\bar {\rm
f}_{\lambda}$ in erg\,s$^{-1}$\,cm$^{-2}$\,$\mu$m$^{-1}$.  The
fractional abundance of $^{12}$CO molecules relative to hydrogen, $\epsilon_{\mathrm{CO}}$,
is assumed to be constant throughout the envelope and it is assumed
that most oxygen is locked-up as CO molecules.  From the literature we
find $\mathrm{f}_\mathrm{CO}=[\mathrm{CO}]/[\mathrm{H_2}]=5\times 10^{-4}$
\citep{knapp:98}, which means that $\epsilon_{\mathrm{CO}}=2.5\times 10^{-4}$.  
We now need to estimate the fraction N$_\mathrm{i}$(CO)/N(CO)
of CO molecules that are excited to
the ($v''=0, J''=2$)-level. We assume the population to be controlled by
radiation transitions between the $v''=0$ and $v''=1$ states. This radiation
is originally supplied by the stellar photosphere at roughly the stellar
effective temperature, $T_\mathrm{eff}$, but diluted by a 
factor of (R$_*/r)^2$ at a distance $r$ from the star. Neglecting 
all loss or addition of photons in the optically thick
spectral lines we find a characteristic temperature of the 
radiation $T_r$ from

\begin{equation}
\label{B}
B(\lambda,T_r) \approx B(\lambda,T_\mathrm{eff}) \times (\mathrm{R}_*/r)^2.
\end{equation}

This estimate gives a radiation temperature at a distance of $3\arcsec$ of
about 340 K, suggesting an $\mathrm{N_{i=2}(CO)}/\mathrm{N(CO)}$-value of approximately 4\%.
This temperature is several times higher than the kinetic temperature
of the gas (the level is super-thermally excited).
Detailed numerical simulations of the radiative transfer in an
envelope model of {\it o} Ceti by Ryde \& Sch\"oier (2000)
verifies the assumption of radiationally controlled populations and
indicates population fractions, $\mathrm{N_{i=2}(CO)}/\mathrm{N(CO)}$, that range from
approximately 3\% to 10\% at distances from $2\arcsec$ to $7\arcsec$. Here, we have adopted a
value $\mathrm{N_{i=2}(CO)}/\mathrm{N(CO)}$ of 5\%, noting that a numerical modelling of the circumstellar
envelope will be necessary for a more detailed and accurate discussion. 
On the assumption that the terminal wind velocity
$v_\mathrm{e}=3\,\mbox{km\,s$^{-1}$}$, 
the mass-loss 
rate deduced
from the measured intensity at 2\arcsec\, away from the star can be
written as

\begin{equation}
\label{mdott}
\dot {\rm M}=0.10\cdot {\rm I_{CO}}(\beta)/\bar {\rm f}_{\lambda}\,\,\,\,\,\,(\mathrm{M_\odot \,yr^{-1}}).
\end{equation}

The observations at $\beta\approx2\arcsec\,$ of the east and west
positions give ${\rm I_{CO}}(\beta)/\bar {\rm f}_{\lambda}=3.1\times
10^{-6}\,\mbox{$\mu$m\,($\arcsec$)$^{-2}$}$ and of the south and north
positions give  ${\rm I_{CO}}(\beta)/\bar {\rm f}_{\lambda}=2.8\times
10^{-6}\,\mbox{$\mu$m\,($\arcsec$)$^{-2}$}$, where $\bar {\rm f}_{\lambda}$ is
estimated from the on-star spectrum.  The mean mass-loss rate found by this
analytic discussion is then $\dot {\rm M}=3.2\times
10^{-7}\,\mbox{M$_\odot$ \,yr$^{-1}$}$ for the east and west positions,
and $\dot {\rm M}=2.8\times 10^{-7}\,\mbox{M$_\odot$ \,yr$^{-1}$}$ for
the north and south positions.  The uncertainty in the derived mass-loss
rate (which is independent of the flux calibration since we use a ratio of
fluxes in the derivation), are due to uncertainties in the position of 
the slit, the level population, the measured scattered intensity, and the 
on-star intensity. Considering these sources of uncertainties we estimate 
a conservative uncertainty in the mass-loss rate of typically a factor of 4.

Our derived value of the mass-loss rate
is in good agreement with values in the literature. For example, \citet{young}
arrive at $(3.6\pm 1.2)\times 10^{-7}\,\mbox{M$_\odot$ \,yr$^{-1}$}$
from the CO($J=3-2$) line for a distance of 100 pc and a
$\epsilon_\mathrm{CO}=3\times 10^{-4}$.  The largest uncertainty in the
estimates of mass-loss rates for miras has until recently been the
distance. The Hipparcos satellite \citep{hipp} has radically improved
this situation. Young's mass-loss rate corresponds to 
$(2.8\pm 0.9)\times 10^{-7}\,\mbox{M$_\odot$ \,yr$^{-1}$}$ with our values of d and 
$\epsilon_{\mathrm{CO}}$. Other estimates of the mass-loss rate lie between
$(1-5)\times 10^{-7}\,\mbox{M$_\odot$ \,yr$^{-1}$}$, see e.g. \citet{danchi:94}, 
\citet{loup}, \citet{planesas:I}, and \citet{knapp:85}.
It should
be mentioned again that the radio profiles are difficult to model due to their asymmetry
and require
an additional component apart from for the standard CSE usually used. There 
may be a bipolar nebula
seen pole-on (cf. the discussion in Knapp \& Morris 1985).

An assumption made in Eq.\,(\ref{bg_formel}) is that the 
circumstellar envelope is {\it tangentially}
optically thin, i.e. along the line-of-sight, at distances from 
the star at which the measurements are made, i.e. 2-7\arcsec.
This optical depth is easily found from

\begin{eqnarray} \label{tang}
{\tau_\mathrm{tang} (p)} & = & \frac{e^2 \lambda ^2}{m_e c^2 m_H}\frac 
{f_{u\leftarrow l} \dot {\rm M} } {\delta \lambda}    
\frac{  \epsilon_{\mathrm{CO}}}{4 \mu  v_e} 
\frac{N_i(CO)}{N(CO)} \times \nonumber\\
 & & \int^{L}_{-L}{\frac{\mathrm{d}l}{p^2+l^2}},
\end{eqnarray}

\noindent
where $p$ is the `impact parameter' which denotes the closest distance from the star at the 
location on the sky of the observations and $l$ is the variable of integration 
(tangential distance along the line-of-sight with the closest point to the star as 
the zero point).
The integration should be calculated over a region of the envelope 
($-\mathrm{L}<l<\mathrm{L}$) where the
differential Doppler shifts are smaller than the width of the line. Approximating 
the line profile
with a step-function, we find this distance to 
be 
\begin{equation}
2 L\approx p \,\frac{\delta \lambda}{\lambda}\frac{c}{v_\mathrm{e}}.
\end{equation}
An estimate of the tangential optical depth of the wind based 
on the derived mass-loss rate 
and the estimated expansion velocity, 
is obtained by solving the integral (Eq.\,\ref{tang}). We find that 
$\tau_\mathrm{tang}\approx7.6\times 10^{15}/p$, where $p$ should be in centimeters. 
Inwards of approximately $4\arcsec\ $ we find that $\tau_\mathrm{tang} (p)
{\small
\raisebox{-0.05cm}{\begin{minipage}{0.2cm} 
\raisebox{-0.1cm}{$> $} \\ 
\raisebox{0.1cm}{$\sim$ }
\end{minipage}} }
\,1$, i.e. that the wind is expected to be optically thick along a line-of-sight.
Although the assumption that the optical depth is smaller than 1 is not fulfilled 
in the inner part of the observed region 
($\beta \approx2-7\arcsec$), the CO emission  declines smoothly as $\beta^{-3}$ all the 
way in, indicating that
the assumption of the wind being optically thin could nevertheless be correct. The dependence 
on angular 
distance of the scattered intensity is not expected to show the same behaviour in an
optically thick and an optically thin wind. For the K{\sc i} scattering in the wind 
of R Scl \citep{bg:97}
the same situation is found. The $\beta^{-3}$-dependence of the intensity could be 
preserved throughout 
the observed parts of the wind if the circumstellar gas were inhomogeneous, which 
probably is a 
reasonable assumption, (cf. the discussion in Gustafsson et al. 1997). This fact could 
lead to a lowering 
of the optical depth if the structures (`clumps') of the wind are of a suitable 
characteristic size and have a low filling factor in the 
line-of-sight.
Note, that the measured 
decline of the intensity could also be achieved if the lines are optically 
thick and the mass-loss rate is not uniform but vary with time. However, this seems 
to require
a fine-tuning of the mass-loss rate with time which does not seem realistic.
A numerical simulation of the 
Mira wind (Ryde \& Sch\"oier 2000) will 
discuss the optical depths in more detail.

In applying Eq.\,(\ref{mdot}) it was assumed that the line radiation at
distance $r$ from the center of the star is represented satisfactorily by the measured
on-star flux $\bar {\rm f}_{\lambda}$, corrected by the distance factor
$(d/r)^2$.  In practise, the observed
on-star spectrum will also contain strong contributions from stellar
light scattered in the inner envelope in non-radial directions,
provided that the gas is not devoid of scatterers there.  The velocity
divergence then brings the photon out of the core of the line profile
at points along the new direction and the chances for it to be
re-scattered are minimal. Therefore, a considerable fraction of the
photons are lost `sideways' from the envelope due to scattering already
in its inner regions, see also the discussion in \citet{ryde:2000}.

\section{CONCLUSIONS}

The successful detection of scattered photospheric light in the
circumstellar CO vibration-rotation lines of Mira has made it
possible to explore the structure of its intermediate circumstellar
envelope (at distances from the star of approximately $100-1000$ stellar radii).

The variation of the
emission with angular distance from the star ($2\arcsec<\beta<7\arcsec$) is found to 
roughly follow
a $\beta^{-3}$ behavior, $\beta$ being the angular distance on the sky.
Our data indicate that the
mass-loss rate, when averaged over about 100 years, was constant over
a period of about 1200 years, assuming that the wind is optically thin along 
the line-of-sight for the measured 
angular distances.

Our observations provide absolute fluxes scattered in the 
circumstellar CO R(1), R(2), and R(3) lines, and provide corresponding line ratios.
Based on a simple analytic model, 
the mass-loss rate is estimated to be
approximately $3\times 10^{-7}\,\mbox{M$_\odot$ \,yr$^{-1}$}$ which,
in view of the uncertainties, is compatible with earlier estimates in the literature.
A numerical modelling of the wind will be necessary for a more detailed and
accurate discussion of the circumstellar envelope. Such a modelling will be presented in a forthcoming paper
\citep{ryde:2000}.

We have also found that the envelope is  
approximately spherically-symmetric to within a factor of two in density. 
Note, however, the asymmetries found by, for example, \citet{planesas:II}.
We have found indications that a similar asymmetry is also present in the 
CO vibration-rotation
emission intensity at a distance of approximately $3\arcsec$ from the star 
in the east direction as compared with 
the west direction. It is interesting to note that the companion of Mira
is located eastward from the star at a position angle of 
$(108.3 \pm 0.1)^o$ \citep{karovska:97}.

\begin{acknowledgements}

The referee is thanked for very valuable comments.
We should also like to thank J. Valenti for assistance during the observations, and 
H. Olofsson for stimulating discussions.  We are grateful to
J.~Barnes, L. Borgonovo, K. Ryde, and  N. Piskunov for generous assistance.
We also owe a dept of thanks to S. H\"ofner and M. Asplund for valuable comments
on the manuscript. 
We acknowledge financial support from the Swedish National Space Board and the Royal Swedish
Academy of Sciences.  

\end{acknowledgements}

\clearpage

\figcaption[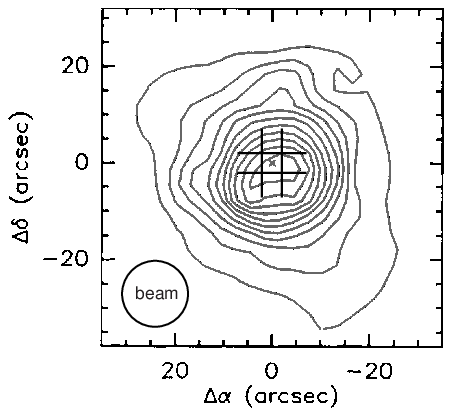]{The slit positions for our observations of 
Mira's envelope. While the spectrograph slit was 30\arcsec\ long, the slit
indicated in the Figure is 14 \arcsec\ long to indicate the regions
in which we detected resonance-scattered CO emission.
The underlying intensity map is taken from \citet{planesas:II}. 
It shows the CO($J=2-1$) line emission
obtained with the IRAM 30 m telescope, in this case for 
$47.7\,\mbox{km\,s$^{-1}$}$, which is close to the stellar 
systemic velocity.  The difference between two
contour lines is $2\,\mbox{K\,km\,s$^{-1}$}$. The IRAM beam is 13
\arcsec\ in diameter. \label{planesas}}

\plotone{planesas.eps}

\clearpage

\figcaption[mira_M_band.eps]{A low-resolution ISO spectrum (R$\approx100$) 
is shown in full
line. The
dashed line indicates the transmission profile of the Johnson M-band
used in the absolute flux calibration of the observations. The width of
the tall box denotes the observed spectral range 
of the Phoenix spectrometer.  \label{Mband}}
\plotone{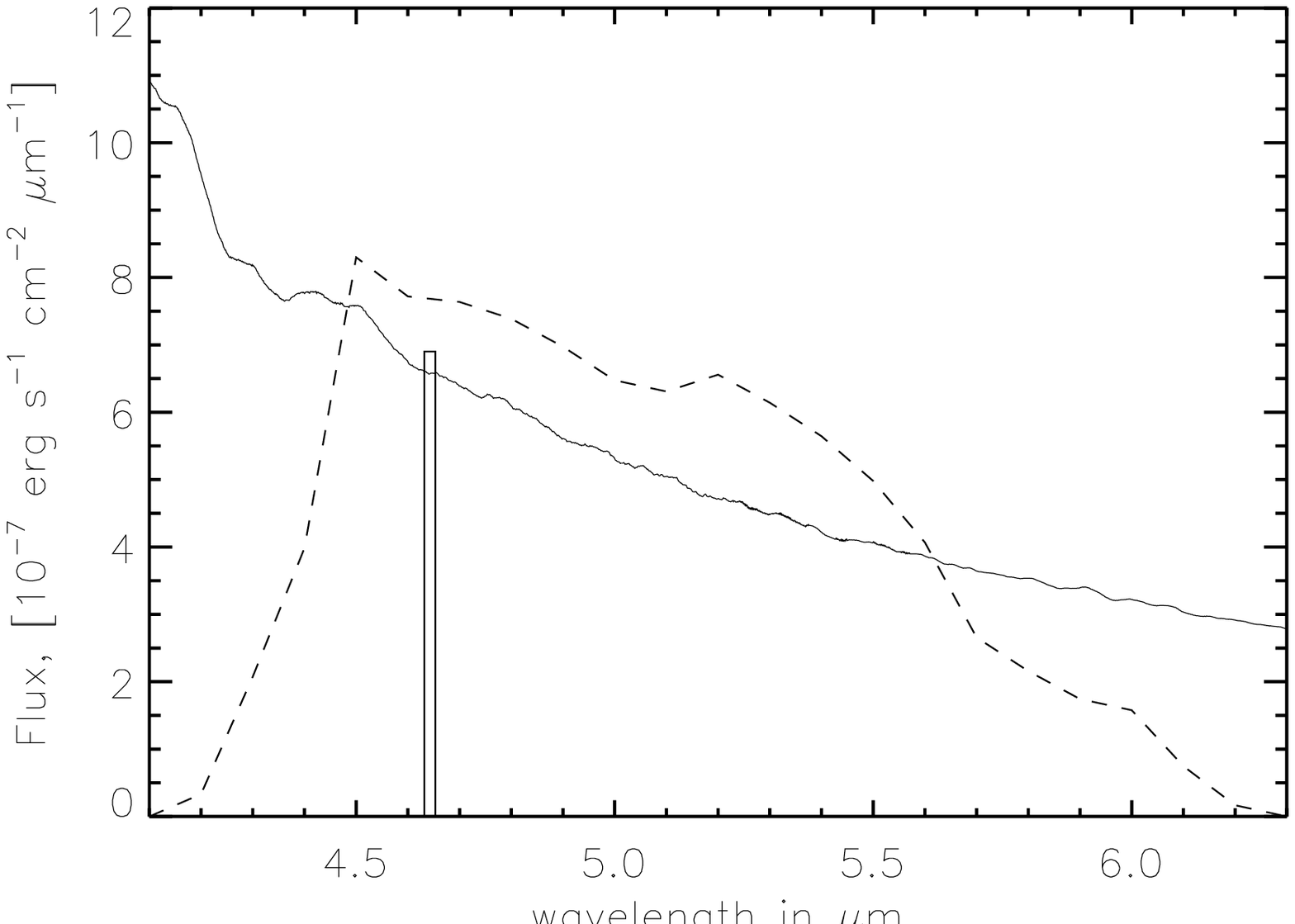}

\clearpage

\figcaption[co_KPNO_oceti_MARCS4.eps] {The observed on-star spectrum of
Mira is shown as a full line. The superimposed dashed line represents
the synthetic spectrum generated from a {\sc os-marcs} model atmosphere
for the stellar parameters T$_\mathrm{eff}=2400\,\mbox{K}$, $\log
g=-1.0$ [cgs], and solar metallicity and assuming a stellar
radius of $3.8\cdot 10^{13}\,\mbox{cm}$ and a distance of
$128\,\mbox{pc}$.  The model fluxes are multiplied by a factor of
1.4 in order to match the observed spectrum. The agreement is
astonishingly good, suggesting that the absolute flux calibration is
reliable.  The dominating spectral features are due to photospheric
water vapor and CO. The telluric water and CO lines are marked ($\oplus$).
Also, the absorption lines due to the photospheric CO 1-0 R(1), R(2), and
R(3) transitions are indicated in the Figure.  Circumstellar absorption
and emission is notable in the cores of the CO lines, especially R(3).
\label{SED}} 
\plotone{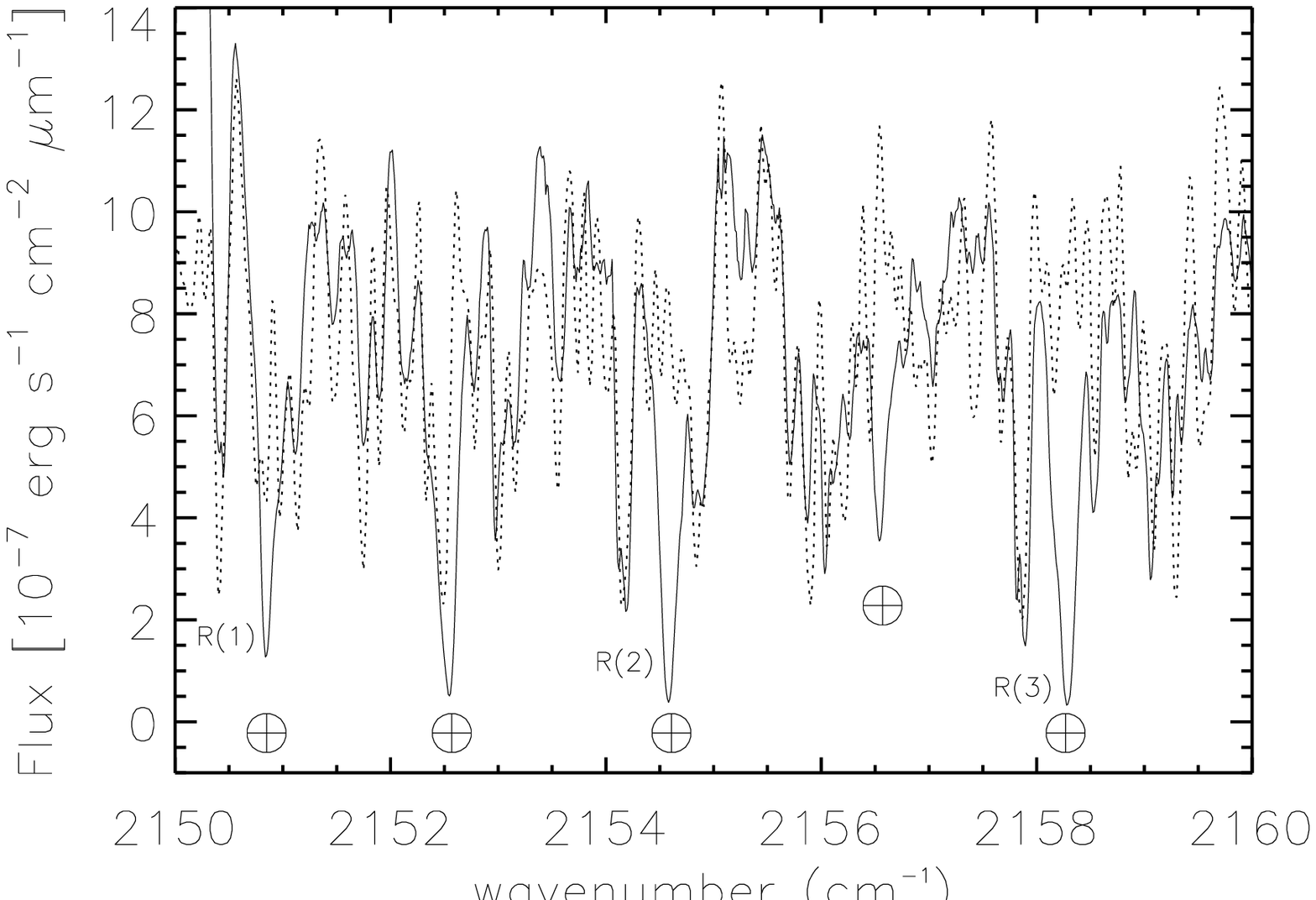}

\clearpage

\figcaption[hastighet2.eps]{The observed scattered spectrum from an 
off-star observation including the R(2) line is
presented as a dotted line.  The solid line indicates a scaled on-star
spectrum. From the Figure we can see that the small emission peak in
the photospheric CO R(2) absorption line is located at the same
wavenumber as the off-star scattered spectrum. This is the result
of some residual circumstellar emission in the on-star spectrum.  
Circumstellar absorption along the line-of-sight through the 
circumstellar shell is present to the right (blue shifted) from 
the emission. The on-star emission and absorption components should nearly cancel
in the measured scattered intensities, which are the differences of our off- and on-star
spectra.
 \label{hastighet}}
\plotone{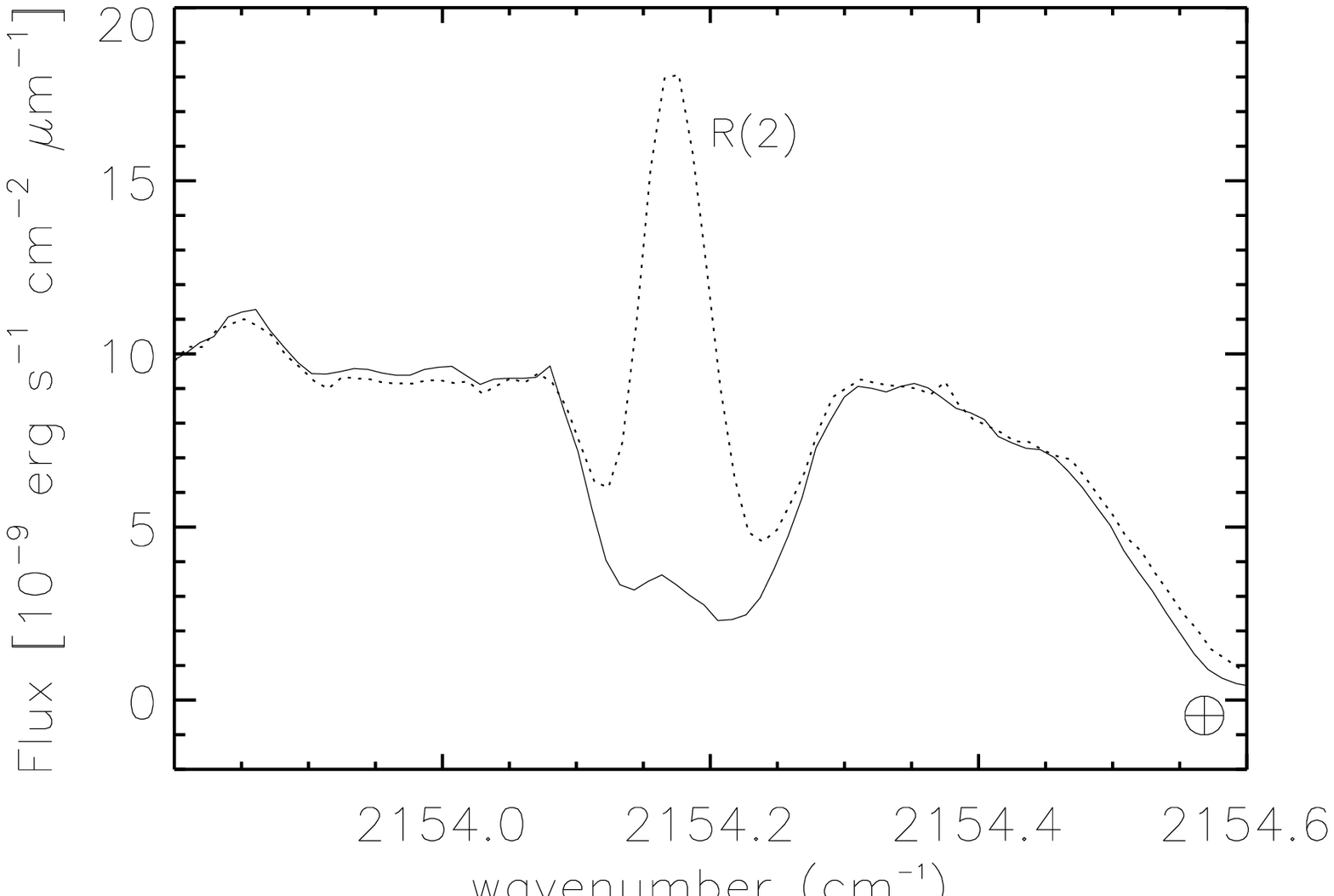}

\clearpage

\figcaption[oceti_off_w4.eps]{The spectra of {\it o}~Ceti (Mira) with
the abscissa showing the wavenumber  scale of the observer.  The
off-star spectrum is shown by a dotted line and the appropriately
scaled on-star spectrum is shown by a solid line.  The three
circumstellar CO vibration-rotation 
emission lines, which are marked  R(1), R(2),
and R(3), `fill in' the corresponding on-star lines which are in
absorption.  To the right (blue) of these lines the telluric CO 
vibration-rotation
R(1), R(2), and R(3) lines are seen. These telluric CO lines and the
two major telluric water vapor lines at $2152.5$ and
$2156.5\,\mbox{cm$^{-1}$}$ are denoted with $\oplus$-symbols.
The stellar CO lines are shifted out of the telluric ones. 
\label{on_off}} 
\plotone{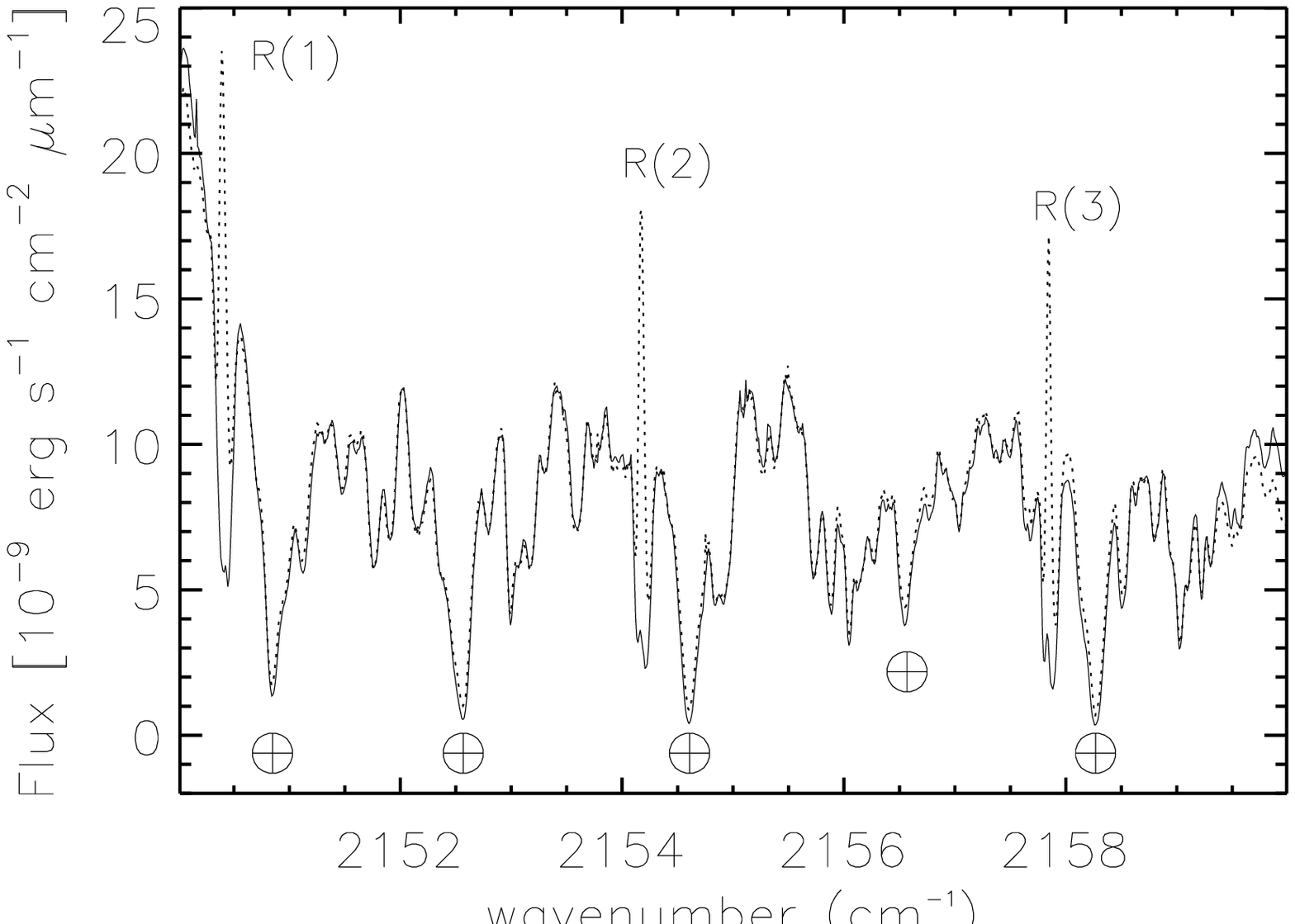}

\clearpage

\figcaption[oceti_em3.eps]{The resulting circumstellar CO 
vibration-rotation {\it
flux} from the wind of {\it o}~Ceti.  The intensity is integrated over
the full slit area. A scaled on-star spectrum is subtracted from the
off-star spectrum. In the four panels the spectra from the observations
with the slit lying tangentially $2\arcsec$ W, E, N, and S,
respectively,  of the star are shown. Note the different ordinate scales.\label{em}}
\plotone{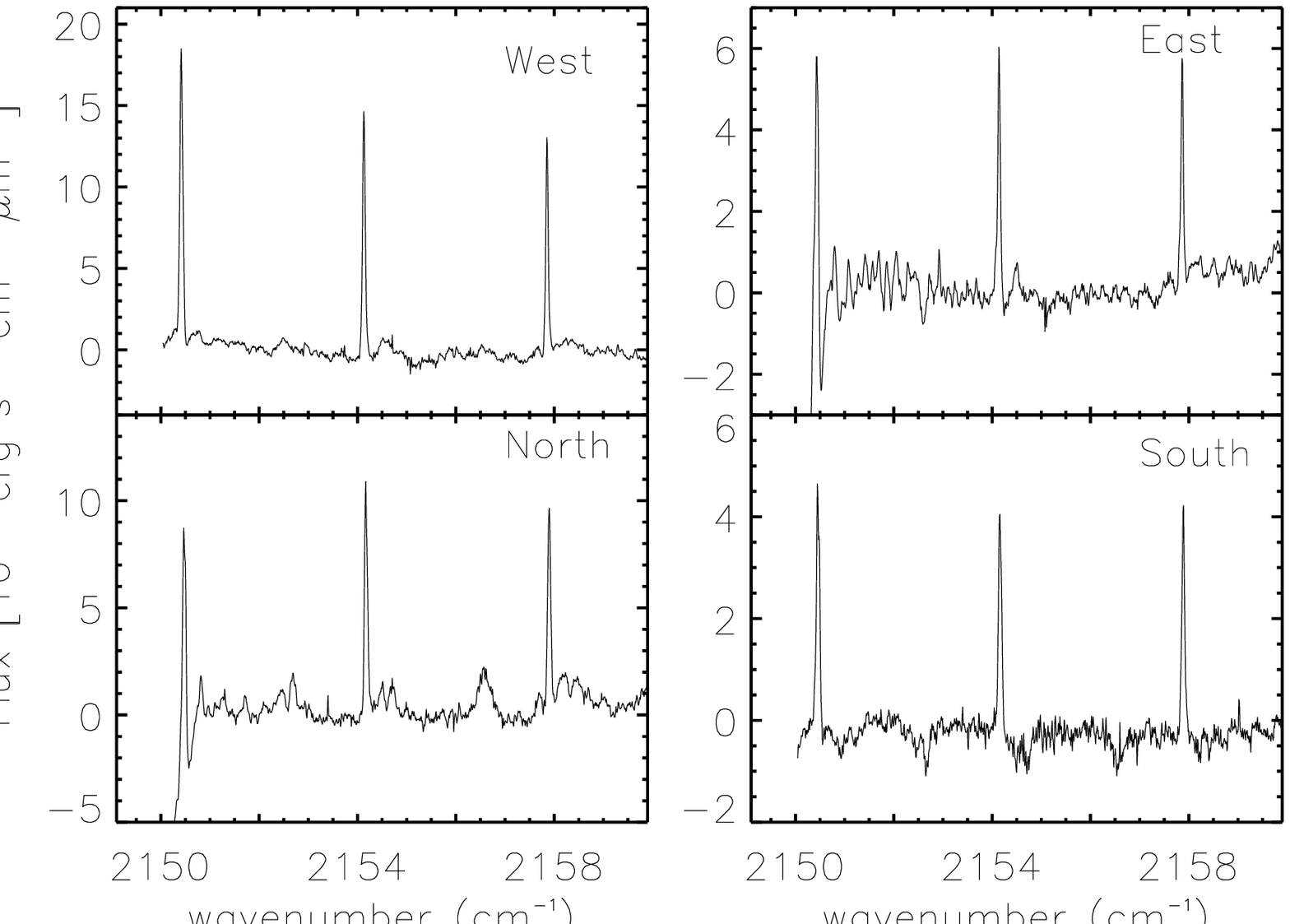}

\clearpage

\figcaption[oceti_SN2.eps]{A set of emission spectra from the west
position is shown to indicate the signal-to-noise ratio at different
distances from the star. The spectra are shifted vertically with
respect to each other.  The top spectrum is measured closest to the
star, i.e. at a distance of 2\arcsec. The subsequent spectra are
measured at 2.2\arcsec, 2.8\arcsec, 3.6\arcsec, and 4.5\arcsec\ away
from the star, respectively.  The intensity scale is given in units of
$10^{-14}\,\mbox{erg s$^{-1}$ cm$^{-2}$ arcsec$^{-2}$ $\mu$m$^{-1}$}$. \label{SN}} 
\plotone{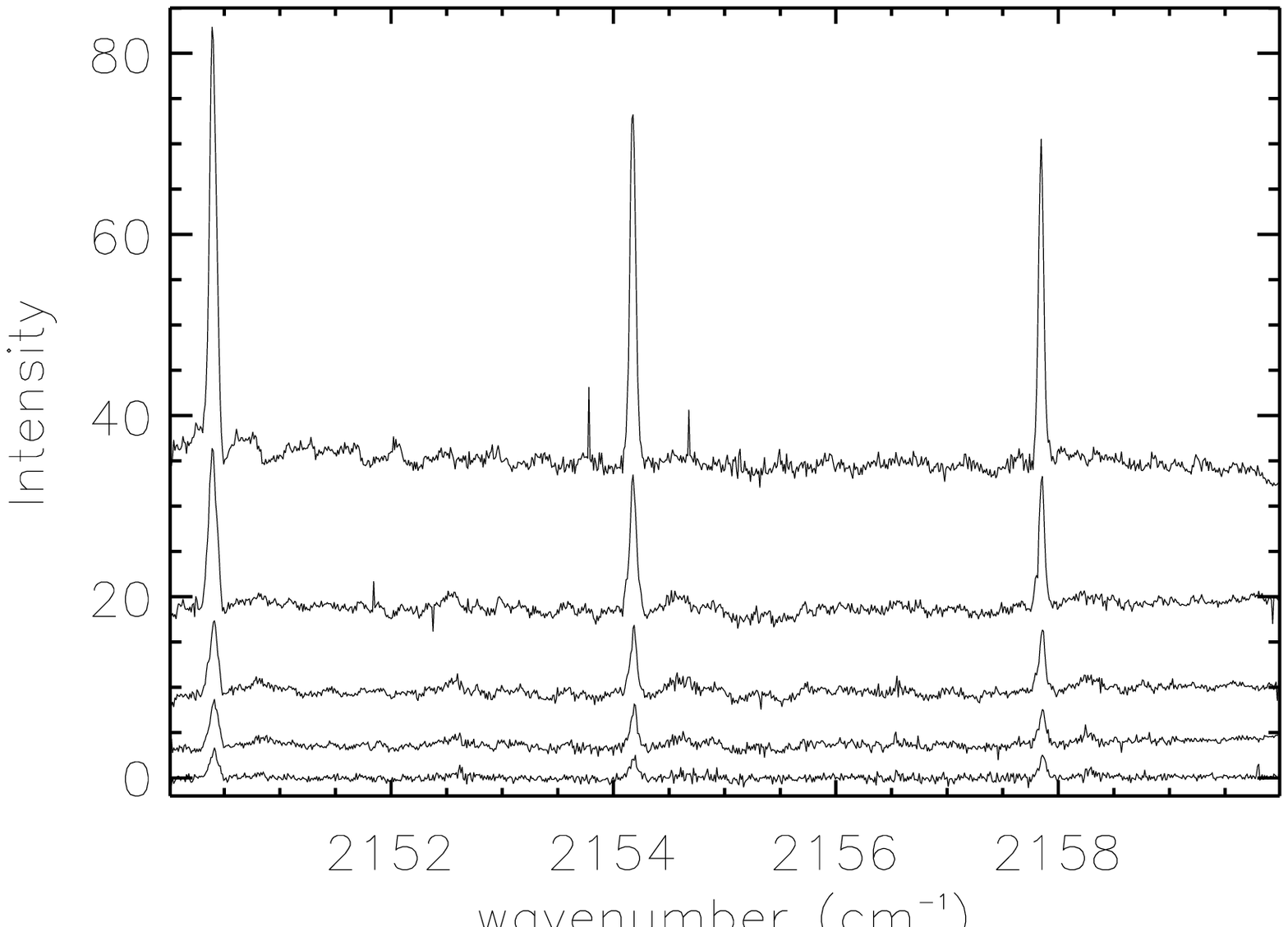}

\clearpage

\figcaption[oceti_beta3_3_000116.eps]{The decline of the {\it
intensity} of the circumstellar emission as a function of angular
distance ($\beta$) from the star. The line is a pure power law
($\mbox{I}\propto \beta^{-3}$). The intensities of
the three vibration-rotation lines [R(1), R(2), and R(3)] are added in order to
increase the signal-to-noise ratio. The two different symbols
represent the two different directions along the long-slit, as
indicated in the different panels.\label{beta3}}
\plotone{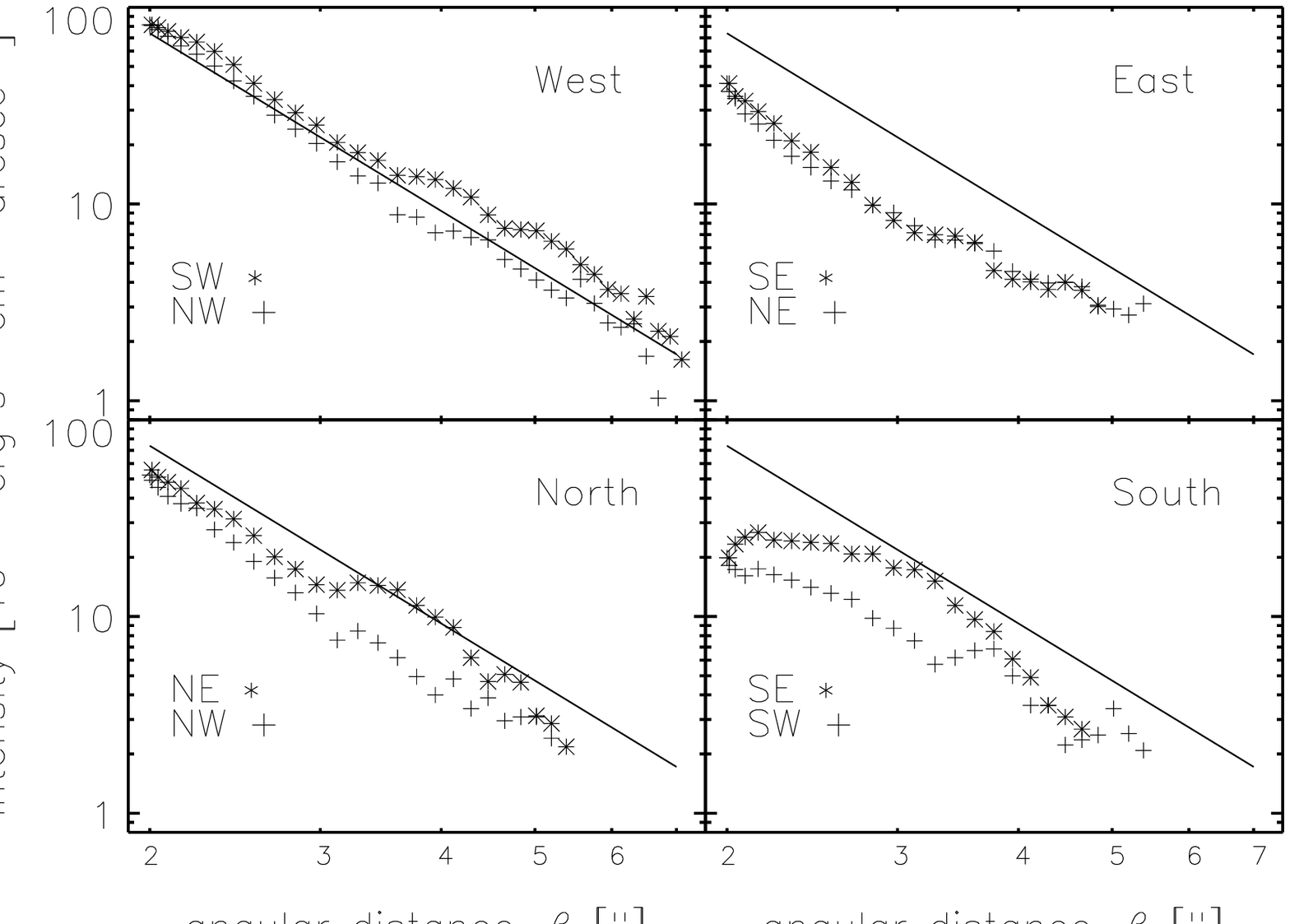}

\clearpage

\figcaption[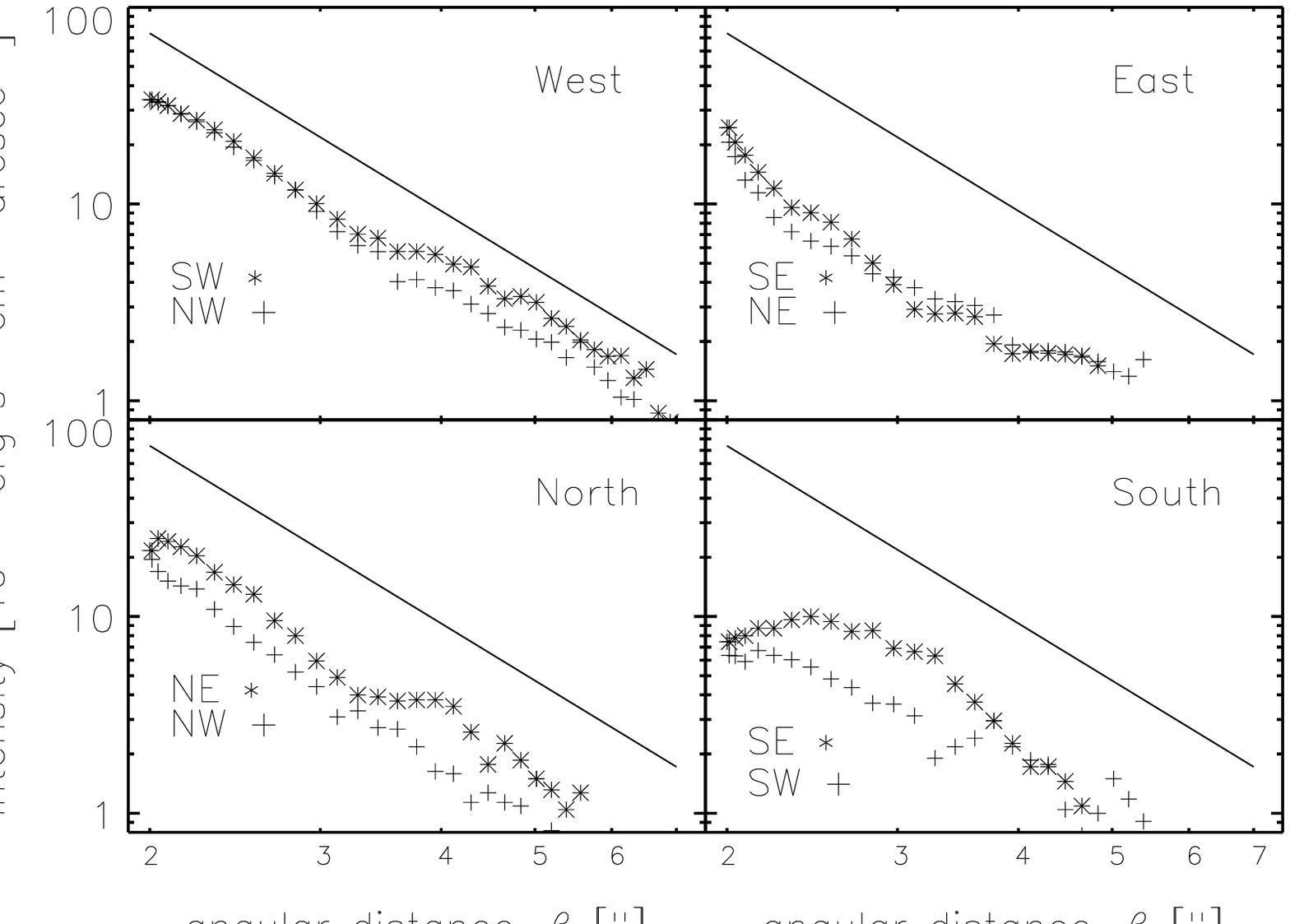]{The decline of the {\it
intensity} of the circumstellar emission as a function of angular
distance ($\beta$) from the star, as in Figure\,\ref{beta3}. Here the scattered light
of the CO R(1) vibration-rotation line is shown.\label{beta3_R1}}
\plotone{oceti_beta3_3_000622_II_R1.eps}

\clearpage
\figcaption[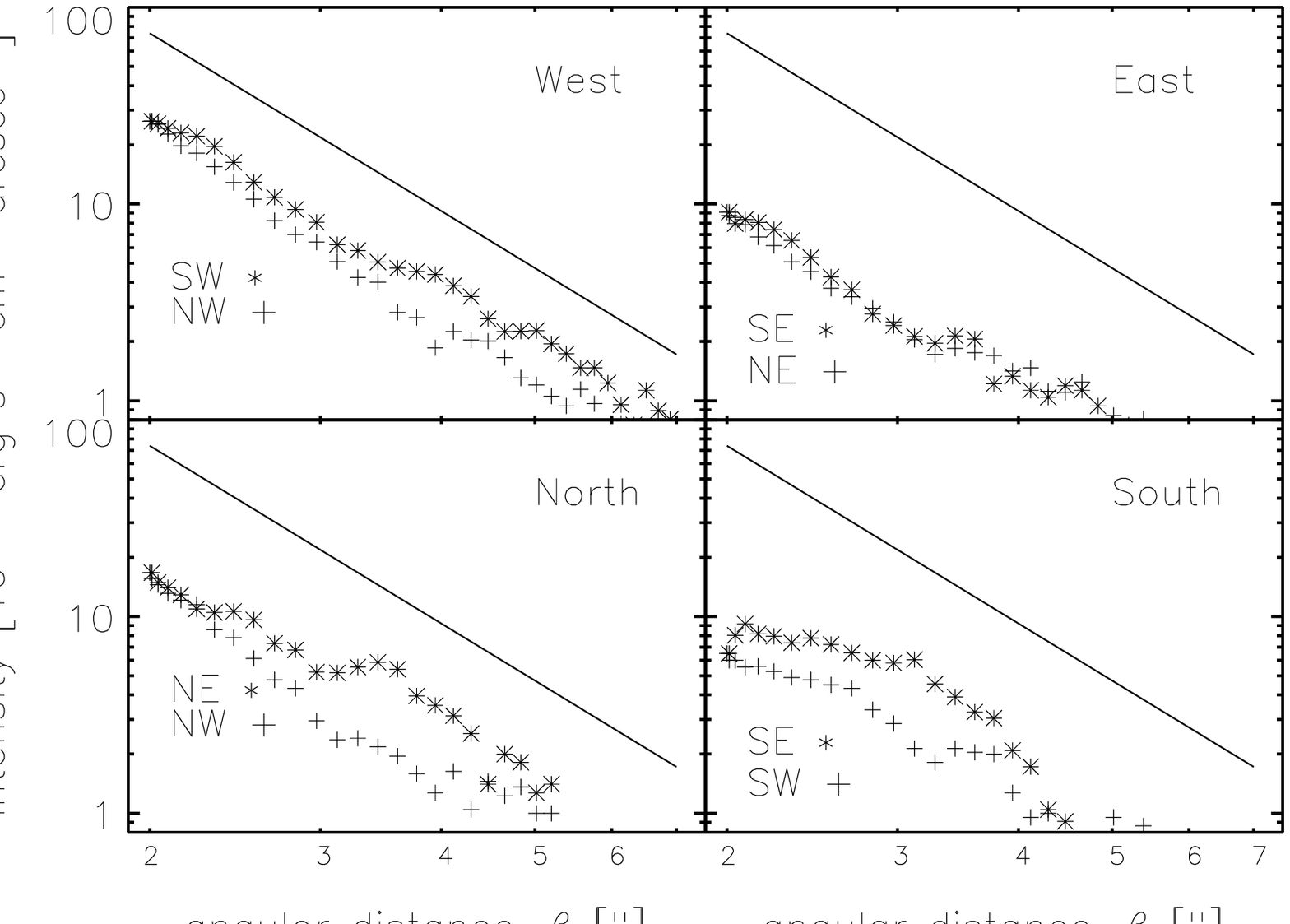]{The decline of the {\it
intensity} of the circumstellar emission as a function of angular
distance ($\beta$) from the star, as in Figure\,\ref{beta3}. Here the scattered light
of the CO R(2) vibration-rotation line is shown.\label{beta3_R2}}
\plotone{oceti_beta3_3_000622_II_R2.eps}

\clearpage
\figcaption[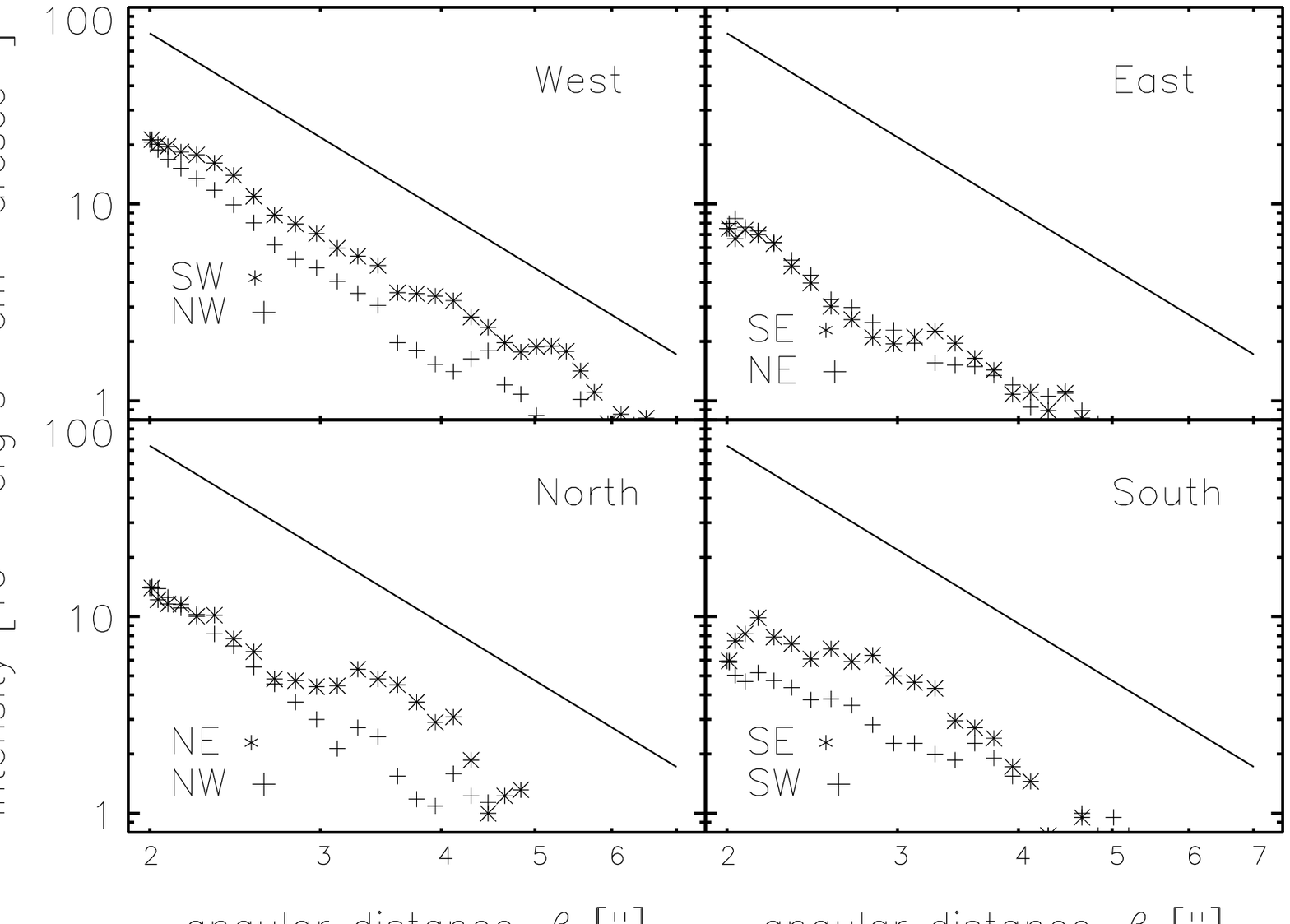]{The decline of the {\it
intensity} of the circumstellar emission as a function of angular
distance ($\beta$) from the star, as in Figure\,\ref{beta3}. Here the scattered light
of the CO R(3) vibration-rotation line is shown.\label{beta3_R3}}
\plotone{oceti_beta3_3_000622_II_R3.eps}

\clearpage

\begin{deluxetable}{ccccc}
\tablewidth{0pt}

\tablecaption{Measured intensities of the CO emission lines, mean
intensity ratios and best-fit slopes \label{kvoter}}
\tablehead{
\colhead{ }                  & \colhead{West}      &
\colhead{East}          & \colhead{North} & \colhead{South}  
}  
\startdata
\sidehead{Intensities [erg\,s$^{-1}$\,cm$^{-2}$\,arcseconds$^{-2}$]:}
  I$_\mathrm{R(1)}$ &  $3.3\times 10^{-12}$ & $2.0\times 10^{-12}$ & $2.3\times 10^{-12}$ & $0.9\times 10^{-12}$\\
  I$_\mathrm{R(2)}$ &  $2.6\times 10^{-12}$ & $0.9\times 10^{-12}$ & $1.5\times 10^{-12}$ & $0.8\times 10^{-12}$\\
  I$_\mathrm{R(3)}$ &  $2.0\times 10^{-12}$ & $0.8\times 10^{-12}$ & $1.3\times 10^{-12}$ & $0.8\times 10^{-12}$\\
\sidehead{Line ratios:}
  R(1)/R(3) &  $1.7\pm 0.2$ & $2.1\pm 0.6$ & $1.7\pm 0.3$& $1.2\pm 0.2$\\
  R(2)/R(3) &  $1.3\pm 0.1$ & $1.2\pm 0.2$ & $1.2\pm 0.2$& $1.1\pm 0.1$\\
\sidehead{Slopes of R(2), $\mathrm d\log I_\mathrm{R(2)} /\mathrm d\log \beta$:}
     &  $-3.3$ (NW) & $-2.5$ (NE) & $-3.0$ (NW) & $-2.3$ (SW) \\
  \noalign{\smallskip}
     &  $-2.8$ (SW) & $-2.5$ (SE) & $-2.7$ (NE) & $-3.2$ (SE) \\
\enddata

\end{deluxetable}

\clearpage

\begin{deluxetable}{cccc}
\tablewidth{0pt}
\tablecaption{Transition probabilities for some CO lines\label{einstein}}
\tablehead{
\colhead{Line ($J''-J'$)}                  & \colhead{wavenumber [cm$^{-1}$]}      &
\colhead{f$_{J''\rightarrow J'}$}          & \colhead{A$_{J''\leftarrow J'}$ }  
}  
\startdata
R(1) (1-2) & $2150.856$ & $6.791\times 10^{-6}$ & $12.57$ \\
R(2) (2-3) & $2154.596$ & $6.127\times 10^{-6}$ & $13.55$ \\
R(3) (3-4) & $2158.300$ & $5.850\times 10^{-6}$ & $14.14$ \\

\enddata

\end{deluxetable}

\clearpage

\end{document}